\documentclass[iop]{emulateapj}
\usepackage{graphicx}
\usepackage{epsfig}
\usepackage{times}

\newcommand{\atlas}{{{ATLAS$^{\mathrm{3D}}$}}}
\newcommand{\afe}{[$\alpha$/Fe]}
\newcommand{\zh}{[Z/H]}

\slugcomment{Received 2014 June 11th; accepted 2014 August 13th}

\shorttitle{Connection between IMF \& stellar populations}
\shortauthors{McDermid et al.}

\begin{document}


\title{Connection between dynamically derived IMF normalisation and stellar population parameters}


\author{Richard M. McDermid\altaffilmark{1,2,*}, Michele Cappellari\altaffilmark{3}, Katherine Alatalo\altaffilmark{4}, Estelle Bayet\altaffilmark{3}, Leo Blitz\altaffilmark{5}, Maxime Bois\altaffilmark{7}, Fr\'ed\'eric Bournaud\altaffilmark{6}, Martin Bureau\altaffilmark{3}, Alison F. Crocker\altaffilmark{8}, Roger L. Davies\altaffilmark{3}, Timothy A. Davis\altaffilmark{9}, P. T. de Zeeuw\altaffilmark{9,10}, Pierre-Alain Duc\altaffilmark{6}, Eric Emsellem\altaffilmark{9,11}, Sadegh Khochfar\altaffilmark{12}, Davor Krajnovi\'c\altaffilmark{13}, Harald Kuntschner\altaffilmark{9}, \\Raffaella Morganti\altaffilmark{14,15}, Thorsten Naab\altaffilmark{16}, Tom Oosterloo\altaffilmark{14,15}, Marc Sarzi\altaffilmark{17}, Nicholas Scott\altaffilmark{18}, Paolo Serra\altaffilmark{14,19}, Anne-Marie Weijmans\altaffilmark{20} and Lisa M. Young\altaffilmark{21,22}}

\affil{
$^{1}$Department of Physics and Astronomy, Macquarie University, Sydney NSW 2109, Australia\\
$^{2}$Australian Gemini Office, Australian Astronomical Observatory, PO Box 915, Sydney NSW 1670, Australia\\
$^{3}$Sub-Department of Astrophysics, Department of Physics, University of Oxford, Denys Wilkinson Building, Keble Road, Oxford, OX1 3RH, UK\\
$^{4}$Infrared Processing and Analysis Center, California Institute of Technology, Pasadena, California 91125, USA \\
$^{5}$Department of Astronomy, Campbell Hall, University of California, Berkeley, CA 94720, USA\\
$^{6}$Laboratoire AIM Paris-Saclay, CEA/IRFU/SAp -- CNRS -- Universit\'e Paris Diderot, 91191 Gif-sur-Yvette Cedex, France\\
$^{7}$Observatoire de Paris, LERMA and CNRS, 61 Av. de l`Observatoire, F-75014 Paris, France\\
$^{8}$Ritter Astrophysical Observatory, University of Toledo, Toledo, OH 43606, USA\\
$^{9}$European Southern Observatory, Karl-Schwarzschild-Str. 2, 85748 Garching, Germany\\
$^{10}$Sterrewacht Leiden, Leiden University, Postbus 9513, 2300 RA Leiden, the Netherlands\\
$^{11}$Universit\'e Lyon 1, Observatoire de Lyon, Centre de Recherche Astrophysique de Lyon and Ecole Normale Sup\'erieure de Lyon, 9 avenue Charles Andr\'e, F-69230 Saint-Genis Laval, France\\
$^{12}$Institute for Astronomy, University of Edinburgh, Royal Observatory, Edinburgh, EH9 3HJ, UK\\
$^{13}$Leibniz-Institut f\"ur Astrophysik Potsdam (AIP), An der Sternwarte 16, D-14482 Potsdam, Germany\\
$^{14}$Netherlands Institute for Radio Astronomy (ASTRON), Postbus 2, 7990 AA Dwingeloo, The Netherlands\\
$^{15}$Kapteyn Astronomical Institute, University of Groningen, Postbus 800, 9700 AV Groningen, The Netherlands\\
$^{16}$Max-Planck-Institut f\"ur Astrophysik, Karl-Schwarzschild-Str. 1, 85741 Garching, Germany\\
$^{17}$Centre for Astrophysics Research, University of Hertfordshire, Hatfield, Herts AL1 9AB, UK\\
$^{18}$Sydney Institute for Astronomy (SIfA), School of Physics, The University of Sydney, NSW 2006, Australia\\
$^{19}$CSIRO Astronomy \& Space Science, PO Box 76, Epping, NSW 1710, Australia\\
$^{20}$School of Physics and Astronomy, University of St Andrews, North Haugh, St Andrews KY16 9SS, UK\\
$^{21}$Physics Department, New Mexico Institute of Mining and Technology, Socorro, NM 87801, USA\\
$^{22}$Academia Sinica Institute of Astronomy \& Astrophysics, PO Box 23-141, Taipei 10617, Taiwan, R.O.C.\\}

\altaffiltext{*}{Email: richard.mcdermid@mq.edu.au}


\begin{abstract}
We report on empirical trends between the dynamically determined stellar initial mass function (IMF) and stellar population properties for a complete, volume-limited sample of 260 early-type galaxies from the \atlas\/ project. We study trends between our dynamically-derived IMF normalisation $\alpha_{\rm dyn}\equiv(M/L)_{\rm stars}/(M/L)_{\rm Salp}$ and absorption line strengths, and interpret these via single stellar population- (SSP-) equivalent ages, abundance ratios (measured as \afe), and total metallicity, \zh. We find that old and alpha-enhanced galaxies tend to have {\em on average} heavier (Salpeter-like) mass normalisation of the IMF, but stellar population does not appear to be a good predictor of the IMF, with a large range of $\alpha_{\rm dyn}$ at a given population parameter. As a result, we find weak $\alpha_{\rm dyn}$-\afe\/ and $\alpha_{\rm dyn}-$Age correlations and no significant $\alpha_{\rm dyn}-$\zh\/ correlation. The observed trends appear significantly weaker than those reported in studies that measure the IMF normalisation via the low-mass star demographics inferred through stellar spectral analysis. 
\end{abstract}





\section{Introduction}

The stellar initial mass function (IMF) of massive galaxies has been the focus of much attention recently, triggered by findings that suggest a non-universal and systematically varying form of the IMF among galaxies in the current day universe \citep{vandokkum10,cappellari12}. These findings differ from studies of resolved stellar systems in and near the Milky Way, which indicate that the IMF has a universal form \citep{kroupa02,bastian10}. Strong evidence for a non-universal IMF has come from the application of various independent techniques, including gravitational lensing \citep{auger10}, stellar dynamical modelling \citep{cappellari12,cappellari13b,tortora13,conroy13}, and spectral synthesis \citep{vandokkum10,vdc11,spiniello12,labarbera13,ferreras13,spiniello14}. With few exceptions \cite[e.g.][]{smith12,smith13,peacock14}, such studies so far agree that, in general, galaxies with higher velocity dispersion require a `heavy' form of the IMF in order to account fully for the measured total mass-to-light ratio. While this general picture is one of qualitative agreement, there is ongoing debate as to what key parameters drive the IMF variations. In particular, \cite{smith14} find a notable discrepancy between methods on a galaxy-by-galaxy basis, and conclude that the dynamically-derived IMF normalisation for 34 objects in Atlas3D does not correlate with [Mg/Fe] after controlling for velocity dispersion.

In this Letter, we report on how the stellar population properties of the complete \atlas\/ sample are related to the dynamically-derived IMF normalisation. This expands the sample of 34 objects of  \cite{smith14} to a total of 212 galaxies, studying age, metallicity and \afe, and spanning a significantly larger range of stellar population parameter values.

\section{Observations and Derived Quantities}


All observations come from the \atlas\/ Survey \citep{cappellari11}, and comprise optical integral-field spectroscopy covering half the stellar light, on average (i.e. one effective radius, $R_e$). The spectral data cubes were spatially integrated to form an effective aperture corresponding to a radius of one eighth of an effective radius, $R_e/8$. Such an aperture is similar to those used in other (generally long-slit or single-fibre spectrograph) studies in the literature, and allows our results to be more directly compared to them. The single stellar population (SSP) models of \cite{schiavon07} were used to measure the SSP-equivalent age, metallicity \zh, and alpha-element abundance \afe\/ \citep[with IMF fixed to that of a unimodal power law of the form: $\zeta(m) \propto m^{-2.35}$,][]{salpeter55} using the chi-squared approach detailed in \cite{mcdermid06}, finding the model from an interpolated grid which simultaneously best approximates our measured H$\beta$, Fe5015 and Mg\,$b$ line indices on the Lick/IDS system \citep{worthey97}. A full presentation of the stellar populations for the \atlas\/ Survey is given in McDermid et al. ({\em in prep}.).

Stellar kinematics were measured from our integral-field spectroscopy using the pixel fitting code pPXF \citep{cappellari04}, and the resulting maps of stellar line-of-sight mean velocity and velocity dispersion were fitted using general anisotropic Jeans models \citep[][]{cappellari08}, in conjunction with multi-Gaussian mass models \citep[][]{emsellem94, cappellari02} from SDSS imaging, which are reported in \cite{scott13}. The stellar mass-to-light ratios $(M/L)_{\rm stars}$ we use here were derived accounting for a standard dark matter halo \citep{nfw96}, and correspond to models `B' as described in detail in \cite{cappellari13a}.

The same integral-field spectroscopy was used to measure the spectroscopic stellar mass-to-light ratio (assuming a Salpeter IMF), $(M/L)_{\rm Salp}$. These were measured independently from the SSP parameters (which also assume a Salpeter IMF), using a regularised pPXF spectral fit of the MIUSCAT stellar population models from \cite{vazdekis12} as templates. To be consistent with the stellar kinematics used in the dynamical modelling, the spectra within an aperture extending to one effective radius were summed and used in the spectral fit. In all, 264 templates were simultaneously fitted to this aperture spectrum, giving the mass-weighted mean mass-to-light ratio for each galaxy taking into account a smooth, non-parametric star-formation history. The resulting r-band $(M/L)_{\rm stars}$ and $(M/L)_{\rm Salp}$ values are given in Table 1 of \cite{cappellari13b}.

Using the same nomenclature as \cite{cappellari13b}, we quantify the mass normalisation of the IMF using the IMF parameter $\alpha_{\rm dyn}\equiv(M/L)_{\rm stars}/(M/L)_{\rm Salp}$, such that $\alpha_{\rm dyn}>1$ implies an IMF `heavier' than Salpeter. For the results that follow, we apply the same sample selection as that paper\footnote[1]{This selection is tabulated in column (6) of the online version of Table 1 in \cite{cappellari13b}.}, excluding objects with H$\beta > 2.3$\,\AA\ within an effective radius, on the basis that they exhibit very strong age gradients, thus breaking the constant mass-to-light ratio assumption of the dynamical modelling approach, and making $(M/L)_{\rm Salp}$ ill-defined. In addition, we exclude the object PGC071531, as the kinematic data were of too poor quality to allow a secure measure of $\alpha_{\rm dyn}$. The following analysis uses the remaining 212 galaxies satisfying these criteria.




\section{Results}

\subsection{IMF and line-strengths} \label{sec:ls}

\begin{figure}
 \begin{center}
  \epsfig{file=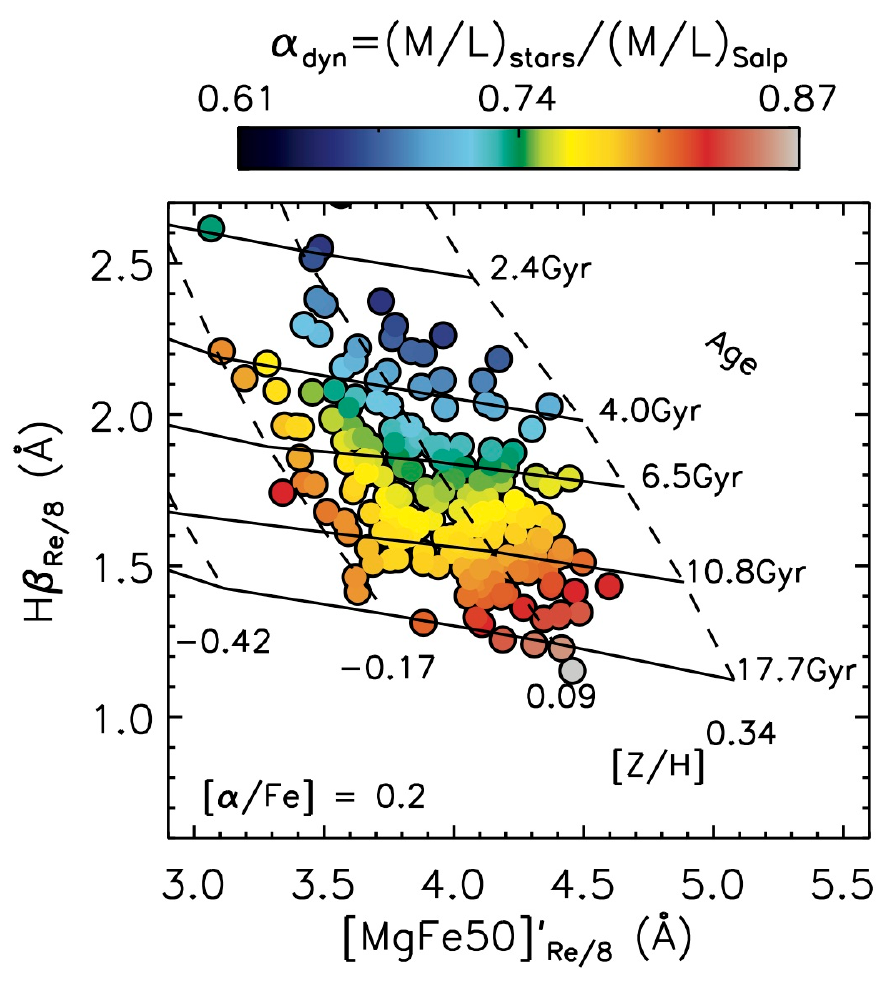, angle=0, width=8.5cm}
  \caption[]{Lick indices H$\beta$ versus the combined index $[$MgFe50$]^\prime$, both measured within an aperture of one eighth of an effective radius. The colour scale indicates the value of the IMF parameter, $\alpha_{\rm dyn}$, defined as the ratio of the mass-to-light ratios derived from dynamical modelling (accounting for dark matter) and stellar population modelling. The measured values of the IMF parameter have been adaptively smoothed using a locally-weighted regression technique to show underlying trends in the distribution. Note that the range of values resulting from this averaging technique is necessarily reduced. A grid of SSP model predictions from \cite{schiavon07} is shown for a super-solar abundance ratio of \afe\/$=0.2$ as indicated in the lower left of the plot. Solid lines indicate lines of constant age; dashed lines constant metallicity, as shown.}
  \label{fig:index_imf}
  \end{center}
\end{figure}

We begin by showing the link between the IMF mass normalisation (via $\alpha_{\rm dyn}$) to simple empirical quantities to make our results independent of stellar population models. Fig. \ref{fig:index_imf} presents the distribution of the IMF parameter plotted in the two-dimensional plane of line strength indices H$\beta$ and $[$MgFe50$]^\prime$\footnote[2]{$[\mathrm{MgFe50}]^\prime = \frac{0.69 \times \mathrm{Mg}b + \mathrm{Fe5015}}{2}$ \citep{kuntschner10}}. The former index is sensitive to stellar age; the latter to metallicity in a way that is not strongly dependent on \afe\ \citep{kuntschner10}. The values of the IMF parameter in this plane have been adaptively smoothed using the two-dimensional Locally Weighted Regression robust technique (dubbed LOESS) of \cite{cleveland88}, as implemented in \cite{cappellari13b}\footnote[3]{Available from purl.org/cappellari/software}. The LOESS-smoothed distribution tries to remove observational errors and intrinsic scatter to estimate the mean values of the underlying galaxy population, thus approximating the mean values one would obtain from simple binning of much larger samples.

Overplotted on these points is a grid of SSP parameter predictions from \cite{schiavon07} for a super-solar \afe\/$=0.2$. There is a trend of larger values of the IMF parameter (corresponding to `heavier' IMFs) towards older, metal-rich objects. The trend, however, is not monotonic, with lower metallicity objects also showing relatively high IMF parameter values, on average.

The LOESS technique averages co-spatial data points, and as with any binning/averaging approach, the absolute range of values is reduced compared to the original data, via reduction of intrinsic and measurement scatter (e.g. compare the colour scale of Fig.\,1 with the intrinsic IMF values implied in Fig.\,2). The LOESS-averaged colour-map plots are useful to uncover the underlying trends, but to quantitatively treat correlations of the actual values, in the following section we analyse the two-dimensional projections of IMF with stellar population parameters individually.

\subsection{IMF and SSP-parameters} \label{sec:ml}

\begin{figure*}
 \begin{center}
  \epsfig{file=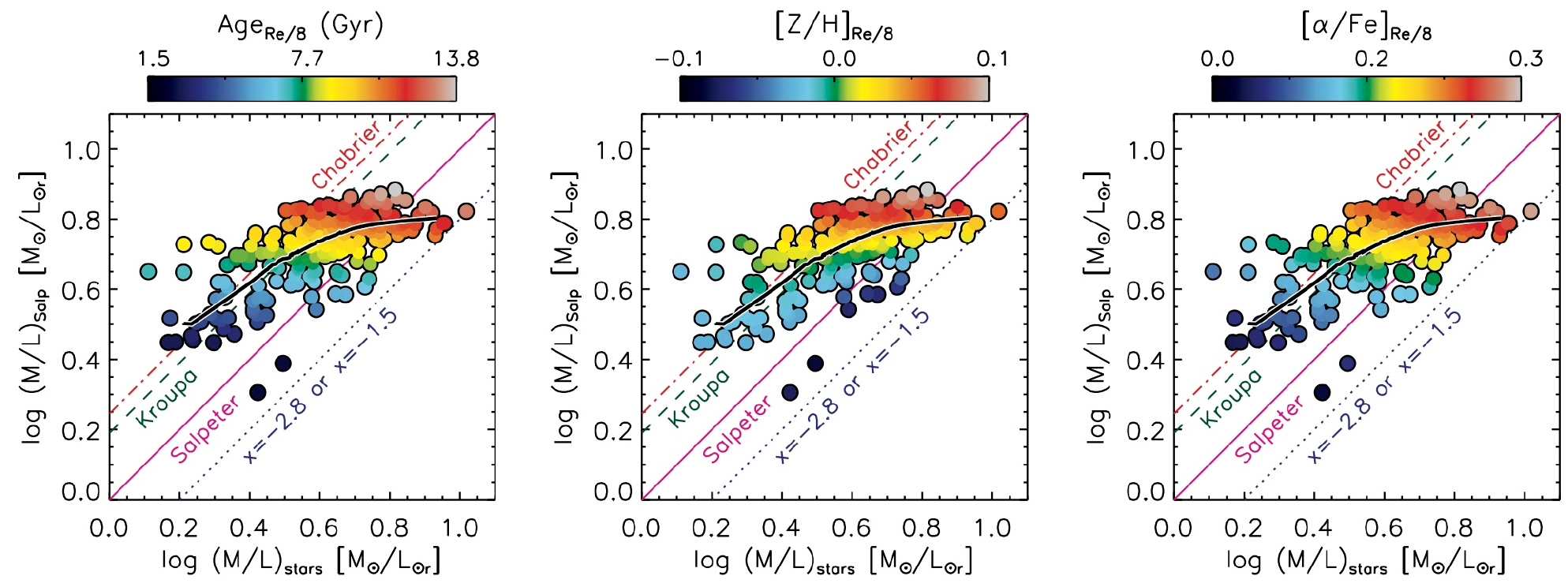, angle=0, width=18cm}
  \caption[]{Each panel compares the mass-to-light ratio measured via stellar kinematics (subscript `stars') and stellar populations (subscript `Salp'). The colour scale indicates the SSP stellar population parameter indicated by the colour bar above each plot. Popular IMF mass normalisations are indicated by diagonal lines with labels. The thick black curve traces the ridge-line of the points via a one-dimensional implementation of the LOESS locally-weighted regression technique.}
  \label{fig:ml_imf}
  \end{center}
\end{figure*}

Each panel in Fig.\,\ref{fig:ml_imf} presents our two mass-to-light ratios $(M/L)_{\rm stars}$ and $(M/L)_{\rm Salp}$ plotted against each other, overplotted with diagonal identity lines to indicate the corresponding IMF normalisation, such that the IMF varies perpendicular to these diagonal lines. The thick black line shows the result of a one-dimensional locally-weighted regression (LOESS) analysis, in order to trace the central ridge-line of the mass-to-light ratio points. The measured SSP parameters are indicated by coloured points, with the parameter name given in the plot title. Again, the SSP parameter values have been smoothed using the two-dimensional LOESS algorithm used in section \ref{sec:ls} in order to show the average trend expected from a larger sample of galaxies. These plots are directly comparable to Fig. 11 of \cite{cappellari13b}, where the colour scale was used to show how velocity dispersion varies in this plane.

It can be seen that the general trends agree with what was inferred above, namely that the IMF becomes systematically `heavier', {\em on average}, for galaxies that are older, more metal rich, and additionally, more enhanced in alpha elements. It is clear from these plots, however, that stellar population is {\em not} a good predictor of $\alpha_{\rm dyn}$. The {\em average} trend appears due to the fact that, while older, or alpha-enhanced, galaxies span a larger range of $\alpha_{\rm dyn}$ values going from Kroupa to heavier than Salpeter, the range of $\alpha_{\rm dyn}$ appears largely limited to Kroupa-like values for galaxies with young ages and low \afe. A similar conclusion can be drawn from Fig. 11 of \cite{cappellari13b}, where galaxies with low velocity dispersion are similarly limited to Kroupa-like IMF normalisation. Metallicity shows a broadly similar trend, however the detailed distribution is different from that with age and \afe, with the iso-metallicity bands running more parallel to the IMF normalisation. Notably, there are objects with low metallicity that require an IMF heavier than Salpeter, as in Fig.\,\ref{fig:index_imf}.

We note that the population parameters vary more tightly with $(M/L)_{\rm Salp}$ than $(M/L)_{\rm stars}$, reflecting the strong dependence of stellar population M/L estimates on stellar population parameters. The dynamical M/L estimate is independent of stellar evolutionary effects, which may explain the broader range of population parameters at fixed $(M/L)_{\rm stars}$.

\begin{figure}
 \begin{center}
  \epsfig{file=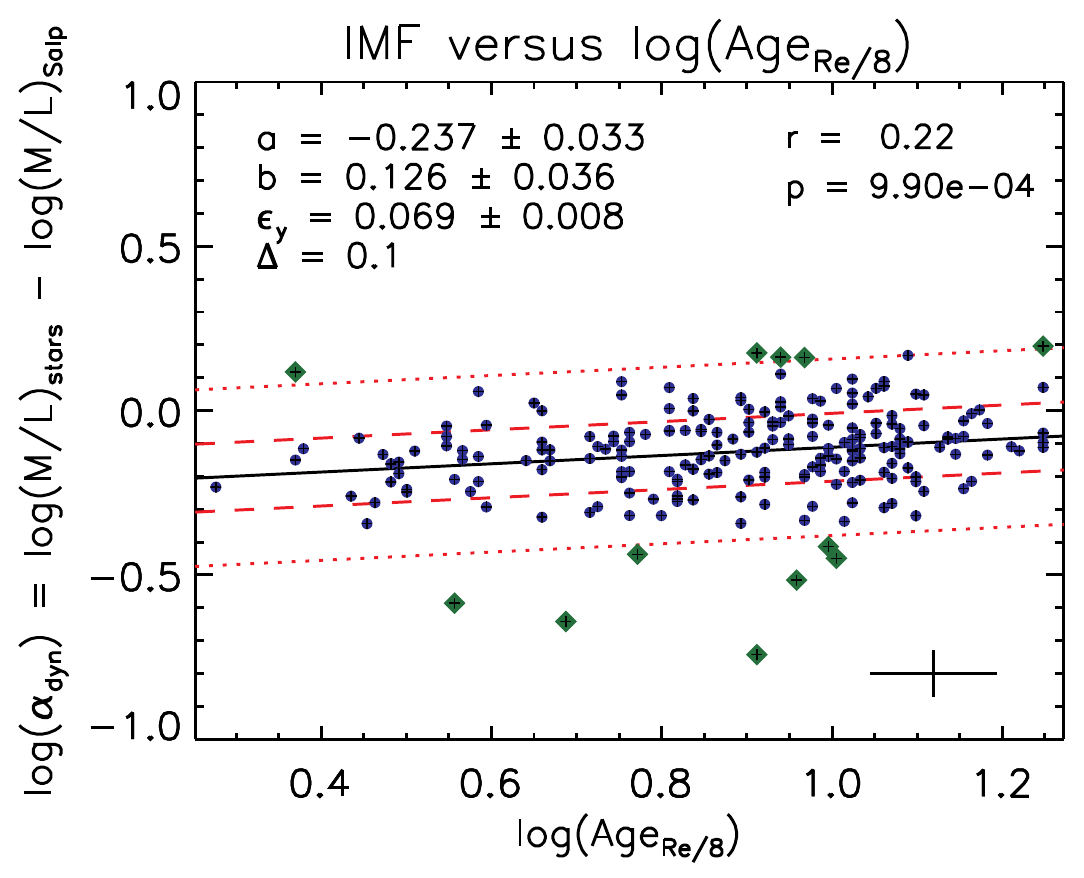, angle=0, width=8.cm}
  \epsfig{file=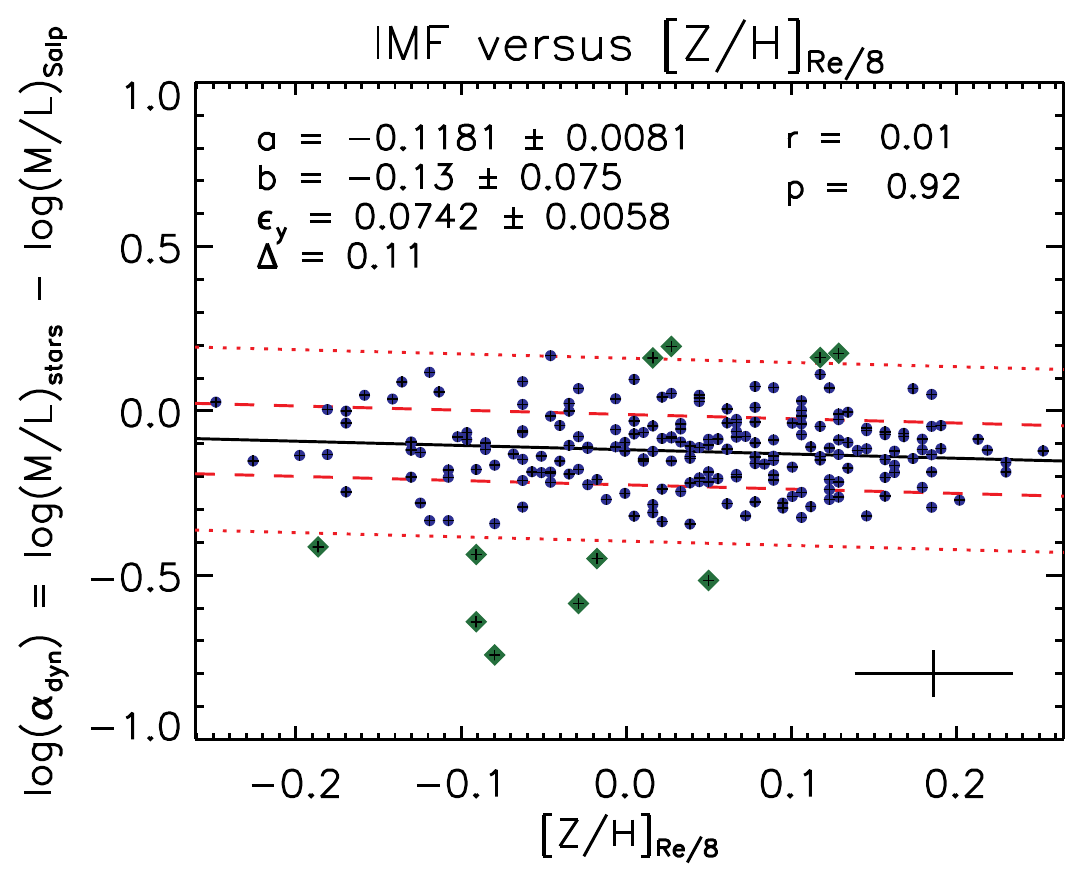, angle=0, width=8.cm}
  \epsfig{file=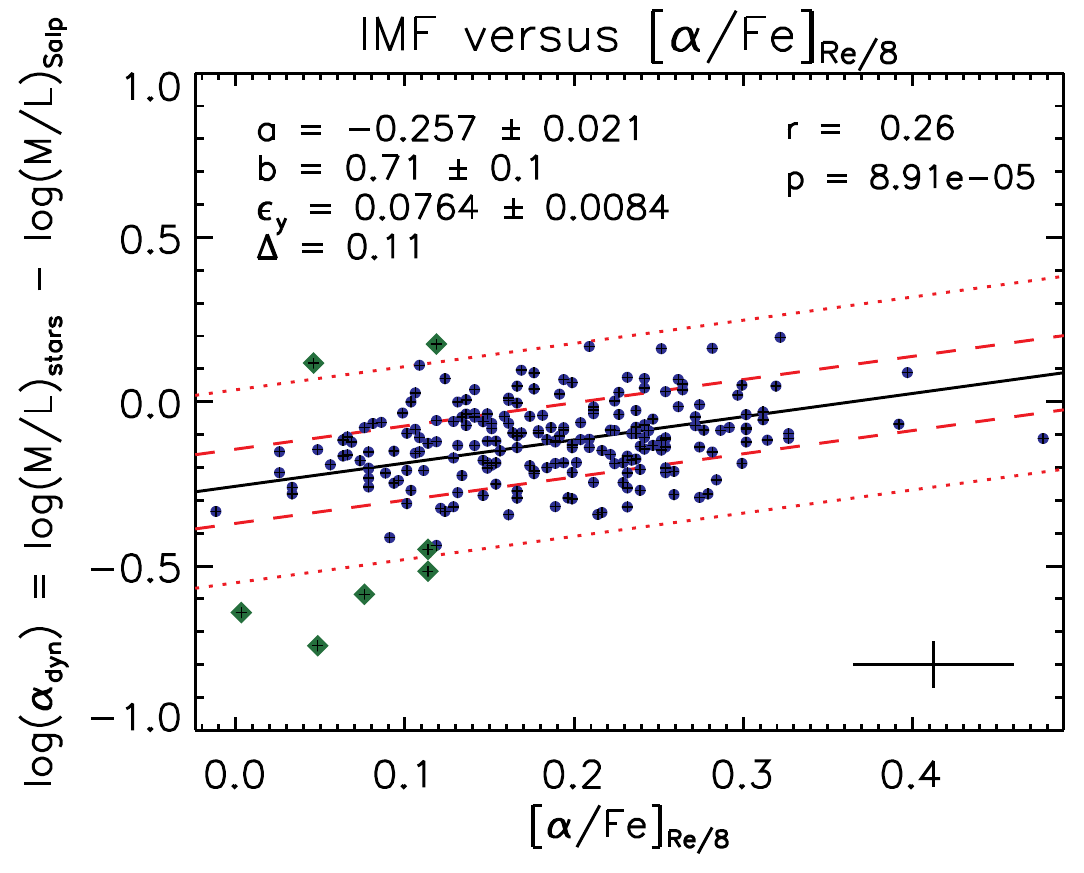, angle=0, width=8.cm}
  \caption[]{Relations between the IMF (vertical axis) and SSP parameters age (top-left), metallicity (top-right) and \afe\/ (bottom). A robust linear fit is shown (black solid line) together with 1$\sigma$ (68\%) and 2.6$\sigma$ (99\%) percentiles as red dashed and dotted lines respectively. Blue circular symbols indicate points included in the fit, with green diamonds showing points rejected during the iterative fitting. Fit parameters are given in the upper left of each panel, giving intercept ($a$), gradient ($b$), intrinsic scatter in the y-direction ($\epsilon_y$) and observed standard deviation around the fit ($\Delta$). Individual errors were included in the fit, but for clarity, the median error for each panel is shown in the bottom right of each panel. Values in the upper right give the Spearman's rank correlation coefficient (r) and significance (p). }
  \label{fig:imf_pop_2d}
  \end{center}
\end{figure}

In Fig.\,\ref{fig:imf_pop_2d} we present the two-dimensional relations of the IMF parameter and population parameters, where points represent our individual measurements. A robust linear fit of the form $y = {\rm a} + {\rm b}x$ is made in each panel using the LTS\_LINEFIT routine described in \cite{cappellari13a}, which combines the Least Trimmed Squares robust technique of \cite{rousseeuw06} into a least-squares fitting algorithm which allows for intrinsic scatter. Fit parameters are given on the upper-left of each panel. Our relations have around 25\%\/ scatter due to observational errors, with 12\%\/ intrinsic scatter. The Spearman's rank correlation coefficient (r) and the two-sided significance of its deviation from zero ($0<p<1$, with smaller values implying higher significance) are given on the upper right. The correlation coefficient values, which are independent from the assumption of a linear relationship, confirm the presence of weak but statistically significant relationships of $\alpha_{\rm dyn}$ with age and \afe, and no evidence of a correlation with total metallicity.

\section{Discussion}
\label{sec:discuss}

The main result of this paper is that the trends between the dynamically-derived IMF normalisation parameter $\alpha_{\rm dyn}$ and stellar population parameters are weak. We find that old or alpha-enhanced galaxies have {\em on average} heavier IMF normalisation, but a heavy IMF is not only found in those galaxies. Old (and high \afe) galaxies span a large range of IMF normalisation going from Kroupa to Salpeter, while young (and low \afe) ones seems to almost exclusively have a Kroupa normalisation. Metallicity shows a less clear distinction, having a broad range of mass normalisations at all metallicity values (Fig. 3, middle panel). \cite{conroy12} also find a weaker relationship between IMF and total metallicity, \zh, concluding, as here, that there is no compelling evidence of a correlation.

We have explored several avenues to verify the weak trends we find. For example, we find fully consistent results using the mass-weighted age and metallicities from the spectral fits used to derive our $(M/L)_{\rm Salp}$ instead of the SSP values, showing that the different models (Schiavon SSP models versus Vazdekis model spectra) and methods (chi-squared fitting for the best SSP versus pPXF spectral fitting of star formation histories for the mass-weighted values) yield consistent results. Using the spectrum integrated within a full effective radius $R_e$ to derive the SSP parameters (instead of $R_e/8$) gives trends equivalent to Fig.\,\ref{fig:imf_pop_2d} within the 1$\sigma$ uncertainties, with comparable scatter, correlation coefficient values and significance (using this larger aperture, the Spearman (r, p) values become $(0.23, 1.1\times 10^{-3})$, $(0.05, 0.5)$ and $(0.22, 2.1\times10^{-3})$ for the equivalent relations of IMF with age, metallicity and \afe\/ respectively). This reassures us that our choice of aperture is not important, and that possible radial variations in the IMF \citep{pastorello14} do not directly affect our conclusions.

It is well established that stellar population parameters correlate positively with velocity dispersion \citep{thomas05,graves09,thomas10}. Given the positive correlation between velocity dispersion and IMF presented in \cite{cappellari13b}, it is tempting to conclude that the trends shown in Fig. \ref{fig:imf_pop_2d} are simply tracing this underlying trend of IMF with velocity dispersion, as argued in \cite{smith14}. However, the tightest correlation of our stellar population parameters with velocity dispersion is with total metallicity (McDermid et al. in prep.). Conversely, metallicity shows {\em no} evidence for any direct, linear correlation with the IMF. This contrary trend of IMF with metallicity compared with age and \afe\/ suggests that our findings are not purely a result of the underlying correlation of IMF with velocity dispersion, as this would give rise to consistent behaviour between all three stellar population parameters, which {\em all} positively correlate with velocity dispersion.

The weak trends we find here are consistent with the similarly weak trend between $\alpha_{\rm dyn}$ and velocity dispersion (${\rm r}=0.36, {\rm p}=5.7\times 10^{-8}$), reported in \cite{cappellari13b}. A comparison of the \atlas\/ velocity dispersion - IMF relation with several published spectral studies \citep{labarbera13,conroy12,treu10,spiniello14} is given in Fig.\,12 of \cite{spiniello14}. While there is overlap between the various spectral studies and \atlas\/ over the range of velocity dispersion in common ($\sigma > 130$km\,s$^{-1}$), the overall relationship for the entire sample is significantly more shallow than that suggested by the spectral studies. Restricting our sample to galaxies with this range of velocity dispersion (where the agreement is best) results in {\em shallower} gradients in the  $\alpha_{\rm dyn}$ - SSP relations presented here, with larger uncertainties (a zero gradient with age and \afe\/ is excluded with only 2$\sigma$ confidence rather than 5$\sigma$), and smaller correlation coefficients (r$<0.16$) with less significance (p$>0.02$), consistent with no significant correlation of $\alpha_{\rm dyn}$ with any population parameter.

Other explanations for the weak trends are possible biases in the dynamical and/or population IMF estimates. An obvious potential systematic problem with the dynamically derived IMF, which has often been invoked in the past, is the treatment of the dark matter content in the dynamical models. We showed in \cite{cappellari12} that for dark matter to explain the IMF trends, it would have to follow the stellar distribution much closer than any current model predicts, making this explanation very unlikely.

Other sources of biases include general problems in the $(M/L)_{\rm Salp}$ derived from stellar population models. Using the SSP models of \cite{vazdekis12} and \cite{BC03} to derive $(M/L)_{\rm Salp}$ from our line-strengths, we find that the weak age trend is removed due to the biased reduction in $(M/L)_{\rm Salp}$ by young populations when using the (luminosity-weighted) SSP approach instead of (mass-weighted) spectral fitting. Any relationship with \zh\/ remains absent using this SSP approach, and the weak trend with \afe, though present, becomes less significant, with best-fit gradients excluding zero at only a 1-4$\sigma$ level. Our weak trends are therefore a general outcome from different models and approaches.

Finally, we note that a direct comparison of the IMF parameters for galaxies in common between our study and \cite{conroy12} yields only a weak relationship between the two studies \citep{smith14}. The relatively small overlapping sample (34 objects), and differences in spatial apertures used in determining the IMF normalisation (their small central aperture versus our IMF analysis using data from, on average, one effective radius) preclude firm conclusions on how this lack of agreement relates to the systematic uncertainties of either study. Removing these remaining issues is the focus of future work.

\section{Conclusions}
\label{sec:conc}

We present the observed trends of stellar population parameters with the dynamically determined IMF for 212 early-type galaxies from the \atlas\/ survey. Using adaptive smoothing to highlight the average trends comparable to studies of larger samples, we find that the IMF normalisation tends {\em on average} to be `heavy' for both weak H$\beta$ {\em and/or} weak metal absorption line strengths. Using single stellar population models, we show that the stellar population properties of age, metallicity and \afe\/ span a broad range of values for any given IMF normalisation.

We present the observed relations of IMF with the SSP properties of the individual galaxies of our sample. We find notably weak relations between the IMF and all three stellar population parameters, with mildly positive correlations of IMF `heaviness' with age and \afe, and {\em no} significant correlation with total metallicity. Taken together with the weak relation between the IMF normalisation and velocity dispersion already presented in \cite{cappellari13b}, these results are somewhat at variance with those from various recent spectroscopic studies. Future efforts for obtaining large samples of individual objects with integral-field spectroscopy that also provides IMF-sensitive spectral features are now critical to permit the responsible systematic uncertainties in local galaxy IMF studies to be resolved.



\acknowledgments

The authors thank the referee for helping to improve this article. The research leading to these results has received funding from the European
Community's Seventh Framework Programme (/FP7/2007-2013/) under grant agreement
No 229517.  MC acknowledges support from a Royal Society University Research Fellowship.
This work was supported by the rolling grants PP/E001114/1 and ST/H002456/1 and visitors grants PPA/V/S/2002/00553, PP/E001564/1 and ST/H504862/1 from the UK Research Councils. RLD acknowledges grants from Christ Church, Oxford and support from the Royal Society Wolfson Merit Award 502011.K502/jd. SK acknowledges support from the Royal Society Joint Projects Grant JP0869822.
TN and MBois acknowledge support from the DFG Cluster of Excellence `Origin and Structure of the Universe'.
MS acknowledges support from a STFC Advanced Fellowship ST/F009186/1.
LY acknowledges support from NSF grant AST-1109803.
The authors acknowledge financial support from ESO.

\clearpage


\end{document}